\documentclass[conference]{IEEEtran}
\IEEEoverridecommandlockouts
\usepackage{cite}
\usepackage{amsmath,amssymb,amsfonts}
\usepackage{algorithmic}
\usepackage{graphicx}
\usepackage{textcomp}
\usepackage{xcolor}
\usepackage{url}
\usepackage[colorinlistoftodos]{todonotes}
\usepackage{amsmath}
\usepackage[normalem]{ulem}
\DeclareMathOperator*{\argmax}{arg\,max}
\DeclareMathOperator*{\argmin}{arg\,min}

\def\BibTeX{{\rm B\kern-.05em{\sc i\kern-.025em b}\kern-.08em
    T\kern-.1667em\lower.7ex\hbox{E}\kern-.125emX}}
\begin{document}

\title{Sensors and Game Synchronization for Data Analysis in eSports}
%\title{Millisecond Synchronization of Sensors and Video for Data Analysis in eSports
%\thanks{The reported study was funded by RFBR according to the research project No. 18-29-22077$\backslash$18.}
%}

\author{\IEEEauthorblockN{Anton Stepanov, Andrey Lange, Nikita Khromov, Alexander Korotin, Evgeny Burnaev, *Andrey Somov}
\IEEEauthorblockA{\textit{Center for Computational and Data-Intensive Science and Engineering (CDISE)} \\
\textit{Skolkovo Institute of Science and Technology}\\
Moscow, Russia \\
*a.somov@skoltech.ru}
}

\IEEEoverridecommandlockouts
%\IEEEpubid{\makebox[\columnwidth]{978-1-7281-2927-3/19/\$31.00 ©2019 IEEE \hfill} %\hspace{\columnsep}\makebox[\columnwidth]{ }}

\maketitle

\IEEEpubidadjcol

\begin{abstract}
eSports industry has greatly progressed within the last decade in terms of audience and fund rising, broadcasting, networking  and hardware. Since the number and quality of professional team has evolved too, there is a reasonable need in improving skills and training process of professional eSports athletes. In this work, we demonstrate a system able to collect heterogeneous data (physiological, environmental, video, telemetry) and guarantying synchronization with 10 ms accuracy. In particular, we demonstrate how to synchronize various sensors and ensure post synchronization, i.e. logged video, a so-called demo file, with the sensors data. Our experimental results achieved on the CS:GO game discipline show up to 3 ms accuracy of the time synchronization of the gaming computer.
\end{abstract}


\section{Introduction}

eSports is organized video gaming where the teams or single players compete against each other with the aim to achieve a specific goal by the end of the game. The eSports industry has progressed a lot within the last decade~\cite{esports-2018}: huge number of professional and amateur teams take part in numerous competitions where the prize pools achieve tens millions US dollars. Its global audience has already reached 380 mln. in 2018 and is expected to reach more than 550 mln. in 2021~\cite{newzoo-2018}. eSports industry includes so far a number of promising directions, e.g. streaming, hardware, game development, connectivity, analytic and training. 

The latter point currently deals with the analytics based on the game statistics and available .demo files allowing for replicating the game and performing fundamental analysis. This kind of analytics is available for both amateur players and professional eSports athletes. However, for gaining deeper knowledge on how to better play and which particular skills must be improved - is the information and analytic services required for professional athletes since the competition is getting harder. Up to date, there are just a few services and research works trying to tackle this problem. At the same time, there is a limited number of research results based on real data, i.e., data collected from the acting professional eSports athletes and in real conditions.
%добавлено из ответа рецензенту, нужно добавить литературу 22-25 в список и перенумеровать ссылки

Psychological studies \cite{welford1980choice}, \cite{brebner1980introduction} states that some individuals have reaction time close to 190 ms (0.19 sec) for light stimuli and about 160 ms for sound stimuli. For correct comparison of reaction time of individual players we should have ability to measure sensors values much faster than human reaction time is. Only in this case we will have acceptable discretization step value. This leads us to $<$ 20 ms requirement for sensors synchronization accuracy.
%конец добавленого

Although the eSports research is in its infancy, there is a fundamental research problem on the data collection synchronization which is relevant for the following areas: Wireless Sensor Networks (WSN) and Body Sensor Networks (BSN)~\cite{wsn-2015, bsn-2015}. It is worth noting that for the most application scenarios the data collected by the WSNs and BSNs are typically homogeneous while the data collected for the eSports are truly heterogeneous: physiological, environmental, video, telemetry (keyboard and mouse), game statistics. 

At present, the problem of heterogeneous data collection could be solved from different points of view. For example, by using a data collection framework~\cite{iot-2014}. However, these frameworks focus primarily on data collection tasks and designing a virtual object (a counterpart of the physical object) instead of solving problems associated with accurate synchronization. In eSports data analysis it is vital to synchronization game video log file, environmental sensors and physiological sensors for getting practically feasible analytics on later steps.

In terms of sensors synchronization, there are two main types of synchronization  \cite{sarvghadi2014overview},  \cite{khediri2012analysis}, \cite{el2016game},  \cite{shahabi2007immersidata}, \cite{lee2008arcade}. First of all, it is \textit{online} sensor synchronization when all sensors share the trigger-sampling signal or all sensors has synced on-board timer. This type of sync is useful since no additional data processing is required after the experiment. Second one is post-synchronization. Some sensors or application, e.g. video games, does not have trigger availability or absolute time stamps. In this case only data processing after the end of experiment is available for synchronization.  

In this work, we propose both synchronization types and share our experience for \textit{heterogeneous} eSports data collection.

This paper is organized as follows: in Section II we introduce the reader to the relevant research work in the area. We present the systems architecture used in this work and its implementation in Section III and Section IV, respectively. Experimental results are demonstrated in Section V. Finally, we provide concluding remarks in Section VI.

%\section{Relevance of content}
%\todo[inline]{Complete this section}
\section{State-of-the-Art}
Recent trends of data analytics in eSports are based on heterogeneous sensing technologies including video. At the same time the eSports disciplines, e.g. CS:GO, are extremely dynamic environments where the game scenario and actions change quickly. For guarantying accurate inference and analytics the data collection has to be properly synchronyzed. It should be noted that for the data collection procedure wireless sensors as well typical 'wired' solutions are involved. Synchronization in WSN is highly relevant topic to our research.

The importance and relevant problems on time synchronization are described in details in \cite{elson2003wireless}. It is clearly stated that physical time synchronization is a crucial point for WSN. There are several popular metrics for evaluating the WSN performance: Accuracy, Computation load, Robustness against failures

In \cite{sarvghadi2014overview}, \cite{sundararaman2005clock}, \cite{rhee2009clock}, \cite{khediri2012analysis}, \cite{sivrikaya2004time} an overview of time synchronization approaches and protocols for WSN was presented. Three clock synchronization frameworks are available: master-slave, peer-to-peer, and distributed. Synchronization is usually done by either aligning the clock readings (called clock offset synchronization, or simply clock synchronization in many references) or aligning the clock rates (called clock rate synchronization, or skew compensation) or both. The so-called drift compensation (aligning the rate of a clock rate) is rarely done \cite{xie2018fast}. 
Selection of an algorithm and a protocol is defined by the design requirements. For all tasks the goal is to achieve high accuracy, low computation load and robustness against many kinds of failure and there are cases to deal with problems of energy cost, energy transfer and environmental issues, i.e., \cite{warier2014spacecraft}, \cite{blaabjerg2006overview}.

For other tasks, e.g. when the biometrical data is collected by the sensors located on different parts of a human body or when the sensors are used as an input tool, the synchronization is an important issue, but reaching high accuracy and low computation load is not essential. For this kind of task, the required energy budget is available, and there are no negative or isolating environmental conditions preventing the data synchronization. We discuss several cases in this section below when few sensors were used to collect and synchronize data acquired from a game and from other sensors, e.g. eye tracker, EEG, EMG, etc., showing the methods to synchronize the data.

For example, in \cite{el2016game} it is suggested to use the microsecond timestamps for all the logged events. It is important to have the same detectable event for each sensor (a fingerprint marked by a mouse and visible by a camera). There is an option to synchronize various sources after the game  session including the situations when external data sources are used, e.g. from psycho-physiological sensors. There is a set of features that are common to all the events that can be logged. A dedicated web server to establish synchronization is acceptable for the analysis of the most game events.

In \cite{mandryk2008physiological} audio streams was captured with a boundary microphone. The game output, the camera recordings, and the screen containing the physiological data were synchronized into a single quadrant video display and recorded onto a hard disk. In this case the video recording was used as a synchronization tool.

ISIS (Immersidata analySIS) system is proposed in~\cite{shahabi2007immersidata}. The idea of ISIS is to collect and synchronized the  telemetry/video data and putting the collected dataset in a context. The system demonstrated reasonable efficiency in the lab trial, but could not handle a large amount of incoming data in real scenario.

Another relevant work is proposed in \cite{lee2008arcade}: the beacons wirelessly communicate with the sink node through the Radio Frequency (RF) signals whereas the sink node is wired to the game-service PC. The beacon is equipped with a wireless sensor node which has an RF transceiver (transmitter and receiver) and an ultrasonic sensor. The node is also embedded in the headband of the game player. The sink node has only an RF transceiver and synchronizes and coordinates the headband and beacons. In \cite{bannach2007waving} the implementation prototype is based on Java-3D for the graphics display and a CRN Toolbox is used for sensor integration.

In \cite{zheng2014multimodal} data from EEG sensor and data from an eye tracking sensor were received by a single device and there was no requirement to solve synchronization task.

This discussion demonstrates that there are no examples when  many sensors are used at the same time. If several sensors are used simultaneously simple basic approaches are used to achieve required quality of synchronization. In eSports data analytics we do need to synchronize multiple heterogeneous data sources and in the next two paragraphs we provide reasoning.

The first example is the analysis of mouse movements by a player during a game. If only mouse coordinates are recorded, then some values would be missing. Data is missed when a player, for instance, rapidly lifts a mouse and moves it to the center of a mouse pad. The only way to understand and analyze the distance between the \textit{before} and \textit{after}  mouse positions is to add a different sensor. For example, it is an option to use an Inertial Measurement Unit (IMU) attached to a mouse or a player's hand, or inside the mouse. An IMU sensor could be connected either by a wire or wirelessly to reduce the interference with the player’s movements. In this case, it is possible to record the data continuously with small time intervals (0.01 s) between the transactions. All mouse movements will be recorded.

The second example deals with the analysis of the in-game actions (in terms of movements) of a player during the game. In a CS:GO environment, a player often performs "jump" or "crouch" actions to overcome the obstacles and avoid opponents. Analysis of game telemetry allows determining the moments where a player made specific movements. However, this analysis does not allow measuring these movements in terms of speed and quality. A player uses fingers to press the keyboard keys to make the  in-game actions. For example, how long the "space" key is pressed during a "jump," or what is an interval between pressing the "space" and "ctrl" keys when the "jump with squat" action is performed.

Recording and analyzing several synchronized sensors at once allows determining the hidden features of a player and helps deal with the situations provided in these two examples.

\section{System architecture}
For ensuring the analysis of a player's behavior, determining the characteristic features and prediction his further actions is possible only if we have enough data generated by heterogeneous sensors. 
%In this case, non-synchronous sensors should not be called a "system."
The appropriate synchronization tool allowing one to achieve high accuracy is mandatory for the data acquisition system.
Synchronization accuracy determined by the hardware capabilities and also by the game played by an eSports athlete. There is a time unit called "tick" in Counter-Strike Global Offence (CS: GO) game discipline. Different game servers have different "tick-rate". Professional players accept the  tick-rate equal to 1/128 s.

From this point of view, we can conclude that the sensor synchronization with 10 ms or less precision is suitable for analyzing players in CS GO.

To get a solution, it's required to perform some tasks: find a stable source of accurate time , select correct synchronization options for a gaming PC, solve issues with synchronization of game telemetry (which often does not have a link to real "off-game" time).

\begin{figure}[htbp]
\centerline{\includegraphics[width=0.45\textwidth]{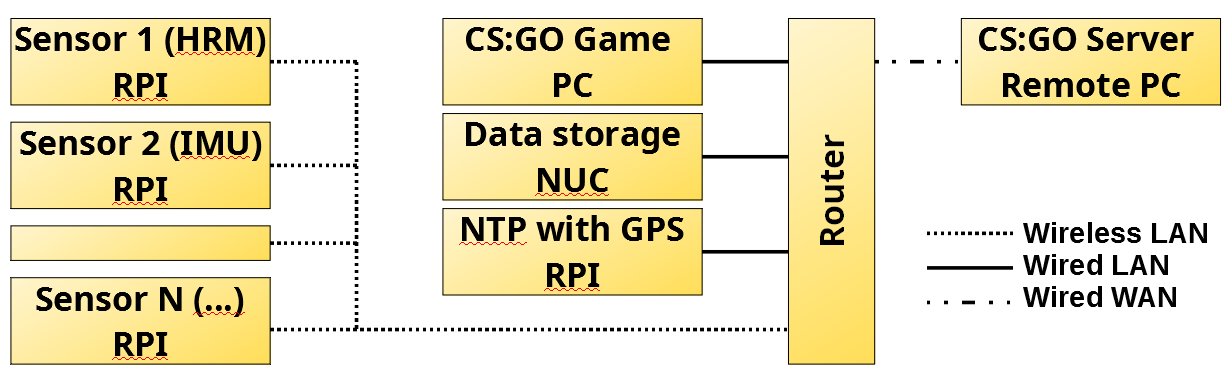}}
\caption{System architecture.}
\label{fig}
\end{figure}

Our system setup is presented in Figure 1. The system has a dedicated storage server (based on Intel NUC PC), gamer PC (high-speed Intel I7 PC with a lot of DDR4 memory and latest GPU card), set of embedded sensors (based on Raspberry PI single-board PC). Also, there is an NTP server with GPS/PPS support. A high-speed wireless router connects devices together. PC with strict requirements to ping value (gamer PC, NTP server) has wired to the router (LAN). Some sensors have wired connection to the game PC: mouse (we call MXY) and keyboard loggers, eye-tracker. Other sensors (heart-rate HRM, Inertial Measurement Unit IMU, skin-resistance GSR, electromyography sensor EMG) have a wireless connection because of placement near an eSports athlete (WLAN). The router has a low latency connection to the Internet (WAN).

\begin{figure}[htbp]
\centerline{\includegraphics[width=0.45\textwidth]{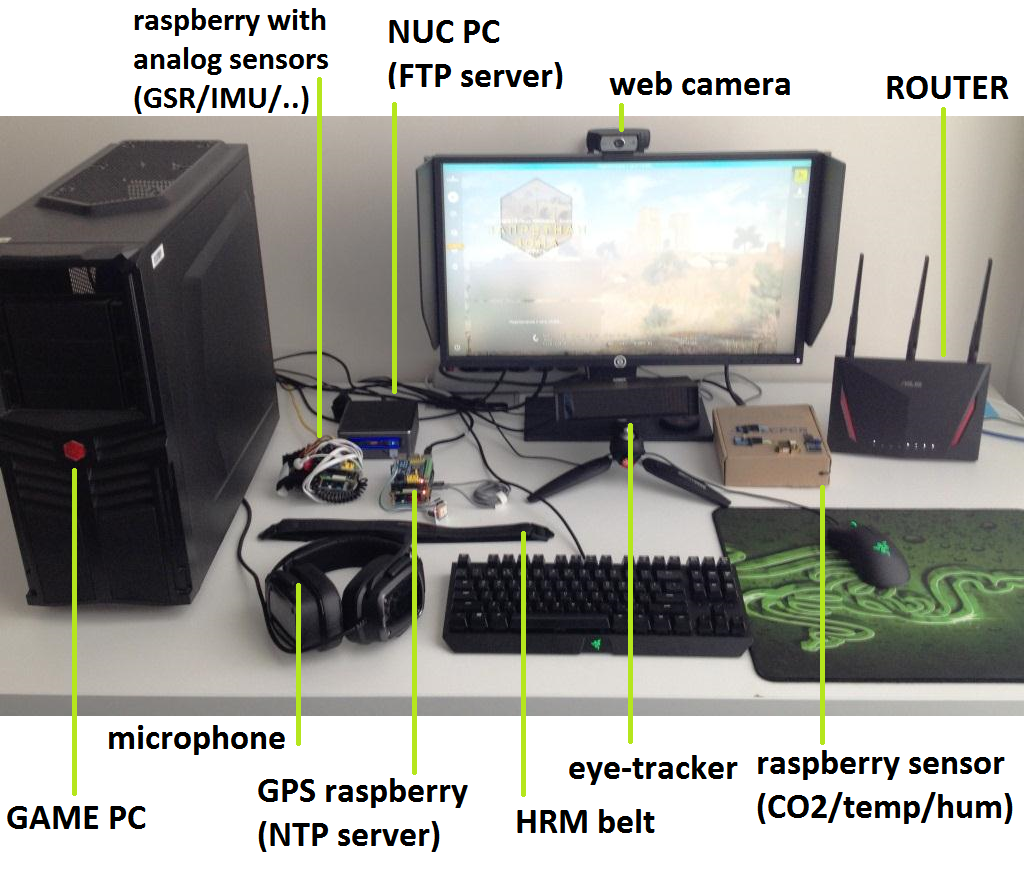}}
\caption{Experimental setup.}
\label{fig}
\end{figure}

Our experimental test-bed is presented on Figure 2. We have set of pro- gamer equipment: mouse, keyboard, headset, monitor with 240fps frame rate, gamer PC (Intel i7/Nvidia 980TI). Also we have high-performance wireless router and separate PC (Intel NUC) for data storage. Our self-made equipment include set of Raspberry PI single board PC. Figure 2 has special marks for RPI based environment sensor (CO2/temp/humidity), RPI based NTP server with external GPS antenna (usually placed near the window) and RPI based analog sensors (HRM/GSR/EMG/IMU).
Typical measurement session does not require any special abilities from player-side. First of all we put necessary sensors on player's body (HRM/GSR/EMG), then activates all recording software and then player plays couple of round on selected game server. All measurements are synchronous by default, except game telemetry which requires post-sync procedure. Details how we able to achieve this task is presented below.

\section{Implementation}
\subsection{Sensors Synchronization}
\subsubsection{NTP Server}
There are many options for building time synchronous systems for industrial applications (one of many examples is  TSN from NI
%\footnote{\url{http://www.ni.com/white-paper/54730/en/}}).
However, the cost of such solutions is considerably higher and these solutions cannot be integrated into the player's PC. It happens because the PC is primarily selected according to the 3D games performance criteria and does not contain specific hardware devices. Therefore, we decided to realize the  synchronization on a single NTP server
%\footnote{\url{https://tools.ietf.org/html/rfc5905}}. 
A reliable server (always available) that would be located close enough (had the minimum delay in transmitting packets over the network) was required. The available options did not suit us by the criteria mentioned above. It led us to the conclusion that the creation of our local time server is required. A single-board computer Raspberry PI 3B
%\footnote{\url{https://www.raspberrypi.org/products/raspberry-pi-3-model-b/}} 
was selected as a server, and a GPS signal was used as a source of accurate time. The signal from the satellite was received by a separate module (based on the MTK MT3333 chipset) having the UART interface and supporting the PPS signal
%\footnote{\url{http://pos.mgb-tech.com/insightpps/}}. 
On Raspbian Stretch OS,
%\footnote{\url{https://downloads.raspberrypi.org/raspbian_lite/images/raspbian_lite-2018-03-14/2018-03-13-raspbian-stretch-lite.zip}}
GPS support packages (gpsd, gpsd -clients, pps-tools) and Chrony time server
%\footnote{\url{https://chrony.tuxfamily.org/comparison.html}}
was installed. Raspberry PI was located near the window for better satellite signal reception and connected to the local area network via a wired interface. The presence of a dedicated PPS signal acquired by a separate IO pin (GPIO) Raspberry PI made it possible to ensure time accuracy in the range of $10^{-5}-10^{-6}$ s (time accuracy of $1 - 10$ us).

\subsubsection{Used Sensors}

Sensors in our system (HRM, IMU, GSR, EMG and others) are deployed on Raspberry PI. Broadcast network "sync" command was sent to  sensors before every measurement. After command reception  special script synchronized local time to local NTP server (Stratum 1) time on each RPI. Feedback status with current time difference was also reported from every RPI to local data storage PC. In this case all RPI were synchronized before measurement starts. Time drift of local RPI time was measured and it has value around $10-20$ ms per hour. In this case sync command was repeated every $10$ minutes. This allows us to have synchronized sensors all the time. 

\subsubsection{Gamer PC Synchronization}

Gamer PC also has several local sensors (mouse and keyboard loggers, eye-tracker and etc.). 
Performing time synchronization on a gaming PC was a separate important task. Players use MS Windows OS on their PCs, which, by default, does not provide accurate time to user (you can check how accurate the clock on the PC is, for example, on a site like \url{www.time.is}). The default settings in Windows 7/8/10 allow you to synchronize time with an NTP server only once a week. At the same time, the average time drift of the clock is 50 ms per hour and even more for an ordinary PC (based on our measurements).
In MS Windows 10 OS build 1607 and newer, there is a way to reduce the synchronization period and get significantly higher time accuracy by setting the registry.
%\footnote{\url{https://docs.microsoft.com/en-us/windows-server/networking/windows-time-service/configuring -systems-for-high-accuracy}}. To do this, first of all you need to make changes to the Windows registry:
%\\1) MinPollInterval to 6 (dec) (Key location HKLM:$\backslash$SYSTEM $\backslash$CurrentControlSet$\backslash$Services$\backslash$W32Time$\backslash$Config)
%\\2) MaxPollInterval to 6 (dec) (Key location HKLM:$\backslash$SYSTEM
%$\backslash$CurrentControlSet$\backslash$Services$\backslash$W32Time$\backslash$Config)
%\\3) UpdateInterval to 100 (dec) (Key location HKLM:$\backslash
%\\$SYSTEM$\backslash$CurrentControlSet$\backslash$Services$\backslash$W32Time$\backslash$Config)
%\\4) SpecialPollInterval to 64 (dec) (Key location HKLM:$\backslash\\$SYSTEM$\backslash$CurrentControlSet$\backslash$Services$\backslash$W32Time$\backslash\\$TimeProviders$\backslash$NtpClient)
%\\5) FrequencyCorrectRate to 2 (dec) (Key location HKLM:$\backslash\\$SYSTEM$\backslash$CurrentControlSet$\backslash$Services$\backslash$W32Time$\backslash$Config)

Then Windows Time Service should be switched to Auto (always loaded after the PC starts) start mode.

An accuracy of the clock within 1 ms require to fulfillment a number of conditions\footnote{\url{https://docs.microsoft.com/en-us/windows-server/networking/windows-time-service/support-boundary}}. In our experiment (taking into account the local time server Stratum 1 based on RPI), all requirements were met with the exception of the ping value (it was $<1$ ms, instead of the required value $<0.1$ ms). However, this did not prevent us to achieve the necessary synchronization accuracy. 

\begin{figure}[htbp]
\centerline{\includegraphics[width=0.45\textwidth]{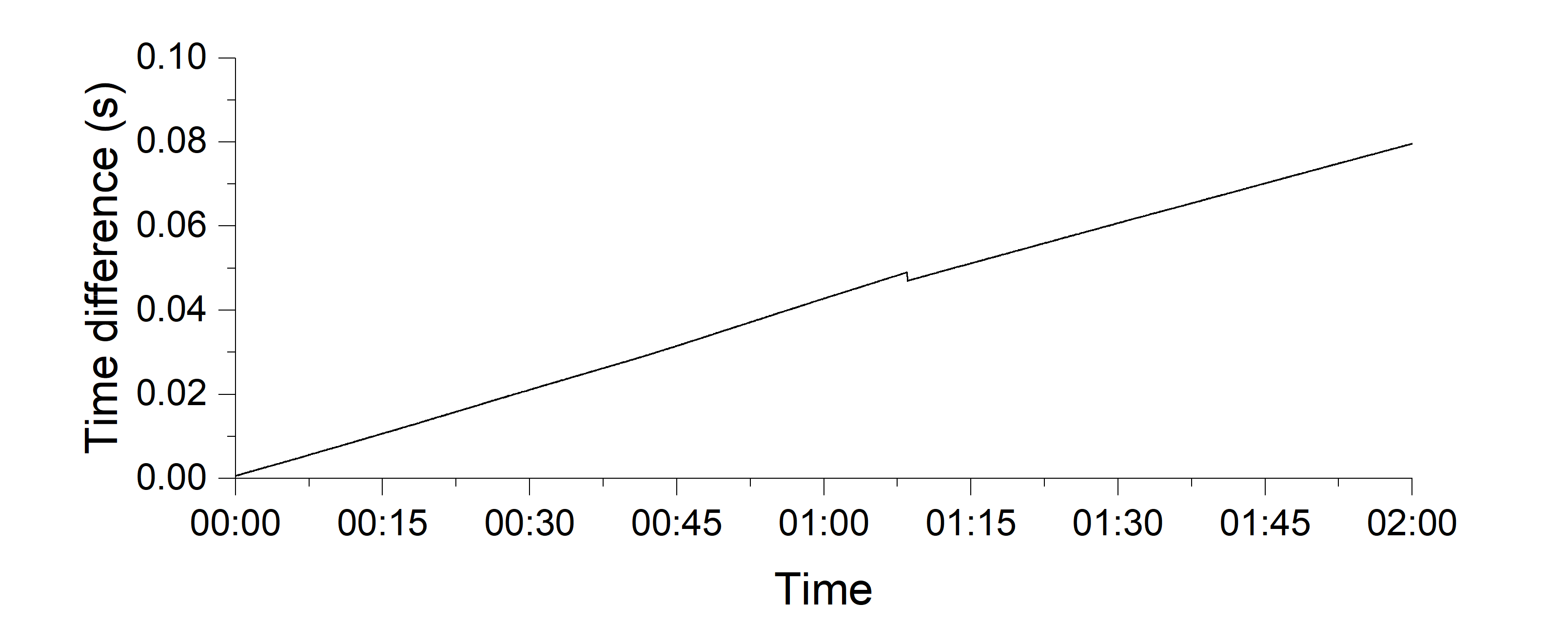}}
\caption{Win 10 PC time accuracy (default settings).}
\label{fig}
\end{figure}

\begin{figure}[htbp]
\centerline{\includegraphics[width=0.45\textwidth]{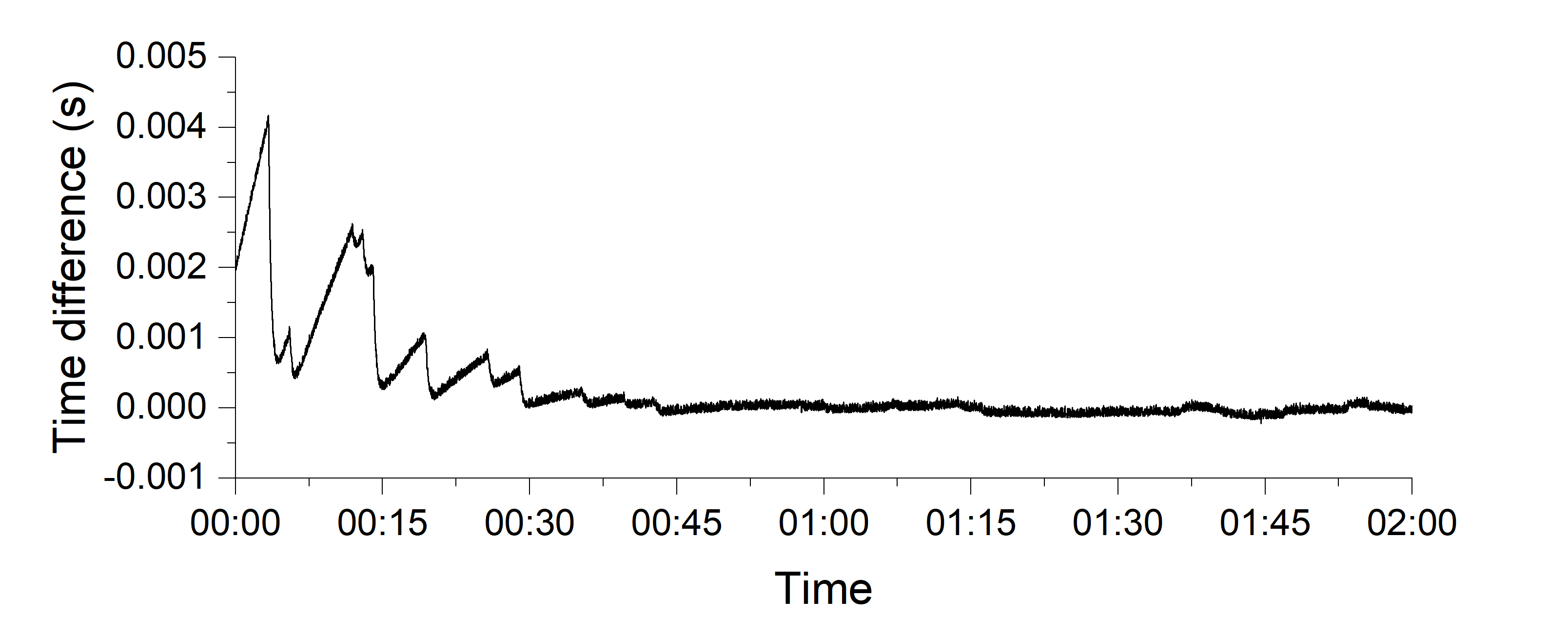}}
\caption{Win 10 PC time accuracy (modified settings).}
\label{fig}
\end{figure}

The figure 3 and 4 shows the clock accuracy on the PC with Win10 (3 is the default settings, 4 is the settings above). The Y axis represents the difference of local time and the time of the time server. The X axis represents the current PC time. It can be seen that with the default registry settings, no time correction is made at all (during several hours), and the clock gradually drifts with the speed $\cong50$ ms per hour. In the case of ''right'' registry settings after some time, drift is compensated by the internal Windows algorithms and the clocks become synchronous with the time server (within $2-3$ ms accuracy).

\subsection{Post synchronization}

\subsubsection{Game Data Synchronization}

Counter-Strike: Global Offensive server uses special bot (GOTV) to log all in-game events into \textit{*.dem} file (replay file). This is a binary file format developed by Valve Corporation
%\footnote{\url{https://www.valvesoftware.com/}}. 
To access \textit{*.dem} file contents we used the official format parser the source code of which is freely available at Valve repository on GitHub\footnote{\url{https://github.com/ValveSoftware/csgo-demoinfo}}.

The parsing tool allows extracting all in-game events (player movements, jumps, shots, death, etc.) from \textit{*.dem} replay into a readable json-like text file. Each extracted event is represented by a dictionary of parameters and their values. Also there is a special time stamp for each event which determines the time when this event happened on the game server.

The game server uses its own discrete time line. Each of these time line moments is called tick. Events are processed by the server (and logged into the \textit{*.dem} file) at the frequency 128 Hz, i.e. there are 128 ticks in each second. The ticks are numbered by integers starting from first tick, which corresponds to the moment when all the game environment is set up on the server.

The \textit{*.dem} format does not have any linkage to the real world time when the replay was recorded (e.g. in UTC time format). This raises the problem of synchronization of the game log with other parts of the system. One may try to restore the UTC time of ticks time using the system \textit{*.dem} file creation time as the time of the last recorded tick in the replay. However, there are no guarantees that the last recorded tick happened at the end of the replay (there may happen several ticks when nothing happened, thus, nothing recorded). Moreover, when the recording of the replay is finished, some time is required to render and save the resulting \textit{*.dem} file. Usually, this time delta is about $1-2$ seconds, which is unacceptably big for synchronization purposes.

We describe an algorithm to perform synchronization of the game log with other sensors used in our system below. Our main idea is to synchronize the event data of the game log with the mouse key pressing of the player, for which we know the exact UTC time. Indeed, from the game log we know all the ticks when player shot from the weapon. At these moment player must have pressed the left mouse button (LMB). Thus, the problem reduces to finding the best match of two binary time series - player fire event from the game and the left mouse button pressing indicator. It is worth noting that not every LMB pressing corresponds to the fire event: players may also use LMB to select element of the game menu, user interface, etc. 

In our system the mouse data is recorded at a rate of 128Hz that is equal to the tickrate of the replay. Thus, for simplicity we can consider the time series of the LMB press indicator to have integer index $t=0,1\dots, N-1$, i.e. the elements are ${g_{0},g_{1}\dots, g_{N-1}\in \{0, 1\}}$, where each time moment $t$ corresponds to some known real UTC time and the gap between two sequential time moments is $\frac{1}{128}$ seconds. The value of $g_{t}$ is $1$ iff the player had the LMB pressed at the time moment $t$, otherwise it is $0$. At the same time we denote the time series of player fire indicator by $f_{0},f_{1},\dots, f_{M-1}$, where $t=1,2,\dots$ is the natural tick index of the game log.

Mathematically, the problem reduces to finding the best integer shift $s^{*}$, such that 
% Надо проверить индексы
\begin{equation}
    s^{*}=\argmax_{s\in[-M+1;N-1]}\bigg[\sum_{m}\mathbb{I}\big[(f_{m}=1)\wedge (g_{m+s}=1) \big]\bigg],
    \label{optimal-shift}
\end{equation}
where for convenience we assume that ${g_{n}\equiv 0}$ for ${n\notin [0, N-1]}$ and ${f_{m}\equiv 0}$ for ${m\notin [0, M-1]}$. In other words, we try to find the time shift, for which we observe the maximal number of matches of in-game player's fire events and the LMB press moments.

Since $f_{m}$ and $g_{n}$ take values in $\{0, 1\}$, it is easy to see that
\begin{equation}
    s^{*}=\argmax_{s\in[-M+1;N-1]}\bigg[\sum_{m}f_{m}g_{m+s}\bigg].
    %\nonumber
\end{equation}
Denote $\tilde{g}_{m}=g_{-m}$ for all $m$. We observe that

\begin{equation}
\sum_{m}f_{m}g_{m+s}=\sum_{m}f_{m}\tilde{g}_{-m-s}=(f\star \tilde{g})_{-s}
\end{equation}

where $\star$ is a discrete convolution operation. Thus, the problem \ref{optimal-shift} reduces to the following:

\begin{equation}
s^{*}=\argmin_{s\in[-M+1;N-1]}(f\star \tilde{g})_{-s}.
\end{equation}

Using the fast algorithm for computing the discrete convolution (see e.g. \cite{burrus1985and}), this problem is solved in ${O\big((M+N)\log{(M+N)}\big)}$ time, i.e. the time complexity of the algorithm is almost linear (omitting the logarithm term) in the total time of replay and mouse data signal.

% A few words why we do not use DTW-like algorithms and that's all.
% Few words in results will be good too)

\section{Results}

During this work we built a sensing system for the eSports athletes monitoring in real-time. This system allows one to measure the synced data from various type of sensors with reasonable time sync accuracy ($<$10 ms). The system details can be helpful not only in eSport field of science, but another field as well, e.g. medical systems.

In this section we discuss how the created synchronous data collection system helps address the problems raised in the introduction, i.e. losses in recording coordinates when moving the mouse and synchronizing game events with recording keystrokes.

In the first case, the presence of two sensors (recording the coordinates of a mouse and an IMU sensor attached to a eSports athlete’s arm) synchronously in time allows data to be recorded even at times when one of them (the mouse) due to its construction (reflection of the laser from the surface of the table or mat) cannot be recorded (mouse transfer in the air). Sufficient synchronization accuracy ($<$10ms) with regard to the measurement period (10ms) allows even at these moments to have a continuous set of data for analysis. An example is shown in the figure 5.

\begin{figure}[htbp]
\centerline{\includegraphics[width=0.5\textwidth]{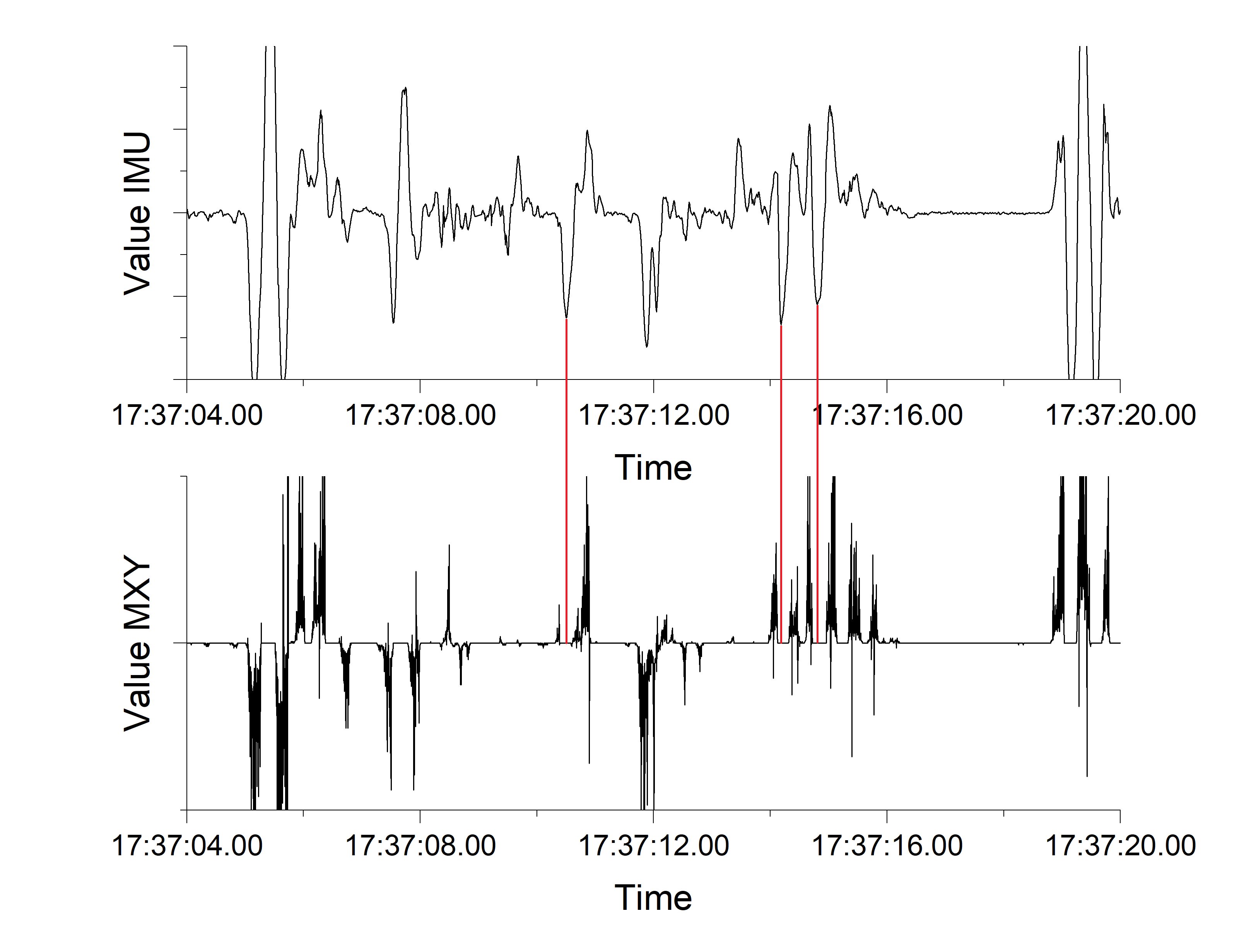}}
\caption{IMU and MXY sensors comparison.}
\label{fig}
\end{figure}

The \textit{Y axis} on Figure 5 represents the sensors value, the \textit{X axis} represents the current time. The upper graph (A) corresponds to the data received from the IMU sensor, the lower graph (B) corresponds of the mouse coordinates data. The graph is shown for a length of $15$ seconds. The red lines marks shows the example situations of missing data in the mouse coordinate record. It is seen that the analysis of two synchronous graphs is more informative, gives less data loss and allows one to detect additional features for data analysis.

\begin{figure}[htbp]
\centerline{\includegraphics[width=0.5\textwidth]{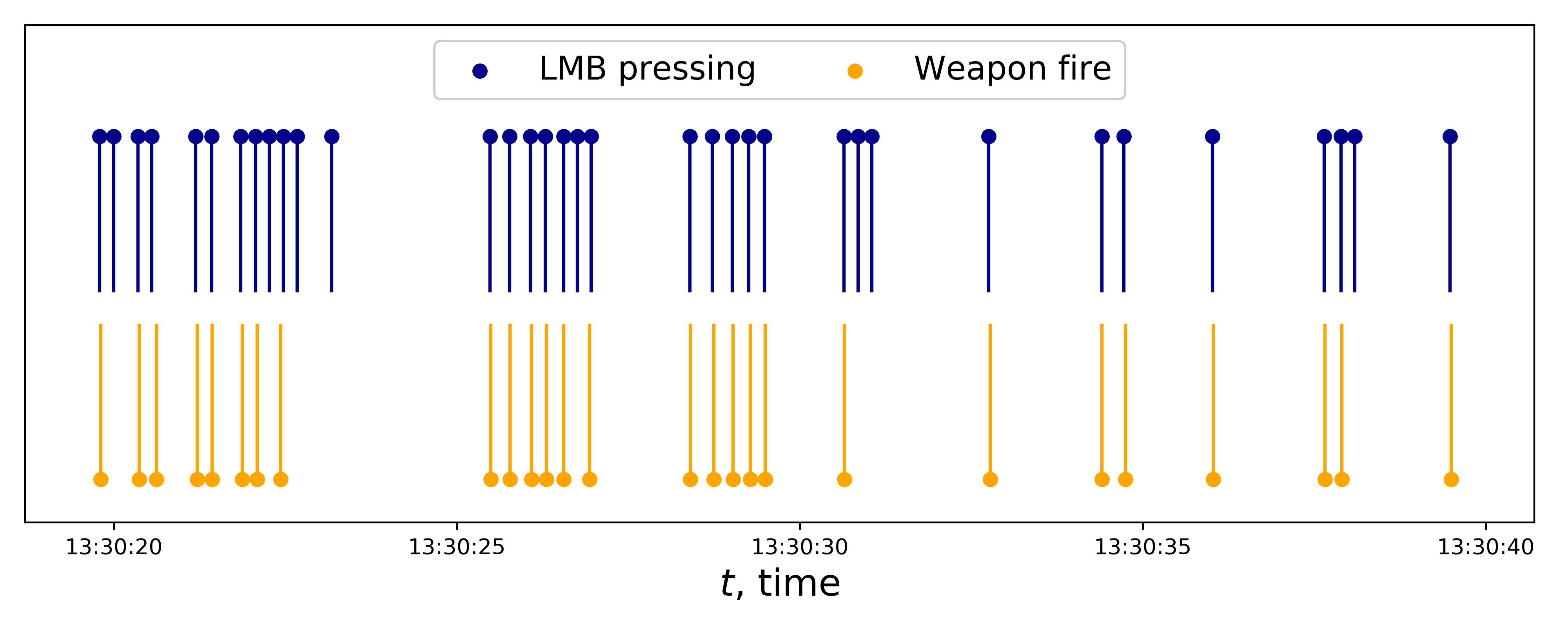}}
\caption{Moments of left mouse button pressings, synchronized with the game events of weapon fire.}
\label{fig:game-sync}
\end{figure}

For the second case (when game events should be synced with keystrokes record) we proposed special post-synchronization algorithm.
In Figure \ref{fig:game-sync} we illustrate the synchronized segments of the game. All weapon fire events match with some mouse log event of left key pressing (the inverse is not true because not every mouse click results in weapon fire).

As we can see in both cases synchronization accuracy $<$10 ms were achieved. This value is $10 - 20$ times smaller than typical humans reaction time. So we can conclude that presented system allows one to collect multi-modal data for future data analysis in the field of physiology, psychology, eSports athletes training and so on.

\section{Conclusion}
During this work a synchronous data collection system was created for monitoring of an esports athlete in CS:GO game. Various sensors were used, including data from a mouse, IMU, keyboard, gaming telemetry, etc. All collected data is  synchronous (with the accuracy of 10ms) due to a configured local Stratum 1 time  server on the Raspberry PI with a GPS signal (and PPS support).
This is a distinctive feature of the created system. The ability to synchronize the time of the gaming computer with an accuracy of $2-3$ ms was confirmed. There is enough accuracy achieved for a comprehensive analysis of the in-game telemetry and physiological indicators of the player. The necessity of using a set of sensors to search for hidden features in the data of eSports athletes is shown. A method for synchronizing CS: GO in-game telemetry with real time is presented.
The proposed approach and methodology could be used to get synchronous data from heterogeneous sensors to ensure high quality of data for further analysis. It's possible to apply them to collecting data from other eSports disciplines , e.g., MOBA, RTS, fighting games, console games,etc. The second possible option is to apply it to a different field, for instance, to medical.

\section*{Acknowledgement}
The reported study was funded by RFBR according to the research project No. 18-29-22077$\backslash$18.
\bibliographystyle{unsrt}
\bibliography{main}

\begin{thebibliography}{10}

\bibitem{esports-2018}
C.~G. Anderson, A.~M. Tsaasan, J.~Reitman, J.~S. Lee, M.~Wu, H.~Steel,
  T.~Turner, and C.~Steinkuehler.
\newblock Understanding esports as a stem career ready curriculum in the wild.
\newblock In {\em 2018 10th International Conference on Virtual Worlds and
  Games for Serious Applications (VS-Games)}, pages 1--6, Sept 2018.

\bibitem{newzoo-2018}
Newzoo.
\newblock Global esports market report, 2018.

\bibitem{welford1980choice}
A~Welford.
\newblock Choice reaction time: Basic concepts.
\newblock {\em Reaction times}, pages 73--128, 1980.

\bibitem{brebner1980introduction}
JMT Brebner.
\newblock Introduction: an historical background sketch.
\newblock {\em Reaction times}, 1980.

\bibitem{wsn-2015}
D.~{Spirjakin}, A.~{Baranov}, A.~{Karelin}, and A.~{Somov}.
\newblock Wireless multi-sensor gas platform for environmental monitoring.
\newblock In {\em 2015 IEEE Workshop on Environmental, Energy, and Structural
  Monitoring Systems (EESMS) Proceedings}, pages 232--237, July 2015.

\bibitem{bsn-2015}
C.~C.~Y. {Poon}, B.~P.~L. {Lo}, M.~R. {Yuce}, A.~{Alomainy}, and Y.~{Hao}.
\newblock Body sensor networks: In the era of big data and beyond.
\newblock {\em IEEE Reviews in Biomedical Engineering}, 8:4--16, 2015.

\bibitem{iot-2014}
S.~{Sasidharan}, A.~{Somov}, A.~R. {Biswas}, and R.~{Giaffreda}.
\newblock Cognitive management framework for internet of things: — a
  prototype implementation.
\newblock In {\em 2014 IEEE World Forum on Internet of Things (WF-IoT)}, pages
  538--543, March 2014.

\bibitem{sarvghadi2014overview}
Mohammad~Ali Sarvghadi and Tat-Chee Wan.
\newblock Overview of time synchronization protocols in wireless sensor
  networks.
\newblock In {\em Electronic Design (ICED), 2014 2nd International Conference
  on}, pages 204--209. IEEE, 2014.

\bibitem{khediri2012analysis}
Salim~el Khediri, Nejah Nasri, Mounir Samet, Anne Wei, and Abdennaceur
  Kachouri.
\newblock Analysis study of time synchronization protocols in wireless sensor
  networks.
\newblock {\em arXiv preprint arXiv:1206.1419}, 2012.

\bibitem{el2016game}
Magy~Seif El-Nasr, Anders Drachen, and Alessandro Canossa.
\newblock {\em Game analytics}.
\newblock Springer, 2013.

\bibitem{shahabi2007immersidata}
Cyrus Shahabi, Kiyoung Yang, Hyunjin Yoon, Albert~A Rizzo, Margaret McLaughlin,
  Tim Marsh, and Minyoung Mun.
\newblock Immersidata analysis: Four case studies.
\newblock {\em Computer}, 40(7):45--52, 2007.

\bibitem{lee2008arcade}
Hyoil Lee, Heekwon Jeong, and JungHyun Han.
\newblock Arcade video game platform built upon multiple sensors.
\newblock In {\em Multisensor Fusion and Integration for Intelligent Systems,
  2008. MFI 2008. IEEE International Conference on}, pages 111--113. IEEE,
  2008.

\bibitem{elson2003wireless}
Jeremy Elson and Kay R{\"o}mer.
\newblock Wireless sensor networks: A new regime for time synchronization.
\newblock {\em ACM SIGCOMM Computer Communication Review}, 33(1):149--154,
  2003.

\bibitem{sundararaman2005clock}
Bharath Sundararaman, Ugo Buy, and Ajay~D Kshemkalyani.
\newblock Clock synchronization for wireless sensor networks: a survey.
\newblock {\em Ad hoc networks}, 3(3):281--323, 2005.

\bibitem{rhee2009clock}
Ill-Keun Rhee, Jaehan Lee, Jangsub Kim, Erchin Serpedin, and Yik-Chung Wu.
\newblock Clock synchronization in wireless sensor networks: An overview.
\newblock {\em Sensors}, 9(1):56--85, 2009.

\bibitem{sivrikaya2004time}
Fikret Sivrikaya and B{\"u}lent Yener.
\newblock Time synchronization in sensor networks: a survey.
\newblock {\em IEEE network}, 18(4):45--50, 2004.

\bibitem{xie2018fast}
Kan Xie, Qianqian Cai, and Minyue Fu.
\newblock A fast clock synchronization algorithm for wireless sensor networks.
\newblock {\em Automatica}, 92:133--142, 2018.

\bibitem{warier2014spacecraft}
Rakesh~R Warier, Arpita Sinha, and Srikant Sukumar.
\newblock Spacecraft attitude synchronization and formation keeping using line
  of sight measurements.
\newblock {\em IFAC Proceedings Volumes}, 47(3):8311--8316, 2014.

\bibitem{blaabjerg2006overview}
Frede Blaabjerg, Remus Teodorescu, Marco Liserre, and Adrian~V Timbus.
\newblock Overview of control and grid synchronization for distributed power
  generation systems.
\newblock {\em IEEE Transactions on industrial electronics}, 53(5):1398--1409,
  2006.

\bibitem{mandryk2008physiological}
Regan~L Mandryk.
\newblock Physiological measures for game evaluation.
\newblock {\em Game usability: Advice from the experts for advancing the player
  experience}, pages 207--235, 2008.

\bibitem{bannach2007waving}
David Bannach, Oliver Amft, Kai~S Kunze, Ernst~A Heinz, Gerhard Tr{\"o}ster,
  and Paul Lukowicz.
\newblock Waving real hand gestures recorded by wearable motion sensors to a
  virtual car and driver in a mixed-reality parking game.
\newblock In {\em CIG}, pages 32--39, 2007.

\bibitem{zheng2014multimodal}
Wei-Long Zheng, Bo-Nan Dong, and Bao-Liang Lu.
\newblock Multimodal emotion recognition using eeg and eye tracking data.
\newblock In {\em Engineering in Medicine and Biology Society (EMBC), 2014 36th
  Annual International Conference of the IEEE}, pages 5040--5043. IEEE, 2014.

\bibitem{burrus1985and}
C~Sidney Burrus and TW~Parks.
\newblock {\em and Convolution Algorithms}.
\newblock Citeseer, 1985.

\end{thebibliography}

\end{document}